\documentclass[12pt]{article}
\usepackage{amssymb, amsfonts, graphicx, hyperref, psfrag, epsfig, mathrsfs, color, amsmath}
\newcommand{\be}{\begin{equation}}
\newcommand{\bea}{\begin{eqnarray}}
\newcommand{\eea}{\end{eqnarray}}
\newcommand{\ba}{\begin{array}}
\newcommand{\ea}{\end{array}}
\newcommand{\ee}{\end{equation}}
\newcommand{\pt}{\partial}
\newcommand{\nn}{\nonumber}

\def\Tr{\mathop{\rm Tr}\nolimits}

\csname @addtoreset\endcsname{equation}{section}
\begin{document}
\begin{titlepage}

  \title{\bf { Entropic force in matrix theory} \vspace{18pt}}
  \vskip.3in

  \author{\normalsize  Wei-shui Xu \vspace{12pt}\\
    {\it\small Department of Physics, Ningbo University}\\ {\it \small Ningbo 315211, People's Republic of China}\\
     {\small E-mail: { \it xuweishui@nbu.edu.cn}} }

  \date{}
  \maketitle

  \voffset -.2in \vskip 2cm \centerline{\bf Abstract } \vskip .4cm
  We consider the entropic force in matrix theory. We find the gravity in bulk can be emergent from the entropic force.

  \vskip 4.0cm \noindent
  \thispagestyle{empty}
\end{titlepage}

\newpage
\section{Introduction}
Recently, Verlinde proposed that the gravitational force in bulk is emergent from an entropic force \cite{Verlinde:2010hp}. When a probe particle approaches to the holographic screen at a small distance  $\Delta x$, it is dissolved into the holographic screen. Finally it becomes a part of the black hole, and increases the entropy of black hole $S+\Delta S$. Then,
\be F=T\frac{\Delta S}{\Delta x} \ee is the so called the entropic force, which will reproduce the gravitational force of this particle experienced in the bulk.

In this paper, we try to give a concrete realization of entropic force in matrix theory. Recently, there were related studies in \cite{Sahakian:2025irl}-\cite{Aldam-Tajima:2026zwq}. We consider $N$ D0-branes in a four-form Ramond-Ramond (RR) background field. Since the Myers effect \cite{Myers:1999ps}, the potential energy for $N$ coincident D0-branes is \bea {\mathcal V}&=&-\frac{T_0}{4\lambda^2}\sum_{a\neq b} \Tr[X^a,
X^b]^2-\frac{i
  T_0}{3\lambda}\Tr(X^iX^jX^k)F^{(4)}_{0ijk}-\frac{T_0}{\lambda}\sum_a\Tr(\Theta^T\gamma_a[X^a,
\Theta]), \label{pot}\\
&&~~~~~ T_0=\frac{1}{g_sl_s},~~\lambda=2\pi l_s^2,~~~~ i, j, k=1,\cdots 3, ~~~~
a=0,\cdots, 9  \nn \eea
where $\Theta$ is a Majorana spinor with 16 components and $X^a$ are
$N\times N$ Hermitian matrices. If the $F^{(4)}$ RR field vanishes, it will reduce to the BFFS matrix model \cite{Banks:1996vh}. In the following, we choose a constant four-form RR field as
\be F^{(4)}_{0ijk}=-\frac{2}{\lambda}f\epsilon_{ijk},~~~~i, j, k=1,\cdots 3.\ee
For the static solution, it satisfies the equation of motion  \be [[X^i, X^j], X^j]+i f\epsilon_{ijk}[X^j, X^k]=0.\ee

Obviously, the commutative $X^i$ is a solution of this equation, then its potential energy vanishes. In addition, there is a spherical
brane solution \be X^i=f J_i,  ~~~i=1,\cdots 3 \label{solution}\ee where $J_i$ are the generators of $SU(2)$ and satisfy the commutative relation $[J_i, J_j]=i\epsilon_{ijk}J_k$. For the
$N$-dimensional irreducible representation, the Casimir operator is \be J^2=\sum_{i=1}^3 J_i^2=\frac{N^2-1}{4}=j(j+1)\ee with the spin $j=\frac{N-1}{2}$ and  $\Tr([J_i,
J_j]^2)=-\frac{N(N^2-1)}{2}$. Thus, the radius of this fuzzy sphere is \be R^2=\sum_{i=1}^3 (X^i)^2=f^2 \frac{N^2-1}{4}=j(j+1)f^2.\ee By inserting this solution (\ref{solution}) into the equation (\ref{pot}), we get the potential energy of this noncommutative solution. It is \be {\mathcal V}^{(0)}=-\frac{T_0 f^4}{24\lambda^2}N(N^2-1).\label{potential}\ee Clearly, the configuration of this fuzzy sphere is more stable than the commutative D0-branes solution. It can prove that this fuzzy sphere is equivalent to a bound state of a spherical D2-brane and $N$ D0-branes in the large $N$ limit \cite{Myers:1999ps}.

To be similar in \cite{Verlinde:2010hp}, we assume this fuzzy sphere to be a holographic screen. In order to realize the entropic force, we add a probe D0-brane to be located at $x^i$. Then the fuzzy sphere with this probe D0-brane is described by $(N+1)\times (N+1)$ matrices, which are
\bea &X^i=\begin{pmatrix}
fJ_i & 0 \\
0 & x^i
\end{pmatrix},~~~X^\alpha=0,\cr
&~~~~~i=1,~ 2,~ 3, ~~~~~\alpha=4, \cdots, 9. \nn\eea
\begin{figure*}[htbp]
 \centering
 \includegraphics[width=0.4\textwidth]{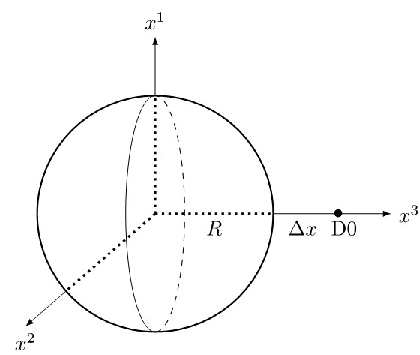}
 \caption{It is a fuzzy sphere with a radius $R$ and a D0-brane at $x^3=R+\Delta x$.}
 \label{fsphere}
\end{figure*}
For simplicity, we assume the probe D0-brane is located at $x^1=x^2=0$ and $x^3=R+\Delta x$ as shown in the Fig.\ref{fsphere}. For this new solution, the potential is same as before (\ref{potential}). Since $\pt\mathcal{V}^{(0)}/\pt{x^3}=0$, there is no interaction between the fuzzy sphere and D0-brane at classical level.

\section{Quantum fluctuations}
Now we consider the quantum fluctuations around this fuzzy sphere. We follow the method in \cite{Aharony:1996bh} and \cite{Gao:2001dp}, and choose the off-diagonal fluctuations as follows \bea && X^i=\begin{pmatrix}
fJ_i & A_i \\
A_i^\dagger & x^i
\end{pmatrix},  X^\alpha=\begin{pmatrix}
0 & A_\alpha \\
A_\alpha^\dagger & 0
\end{pmatrix}, \cr
&&\Theta=\begin{pmatrix}
0 & \theta \\
0 & 0
\end{pmatrix},
~~~~ \Theta^T=\begin{pmatrix}
0 & 0 \\
\theta^T & 0
\end{pmatrix}, \label{fluctuation}\\
&&~~~~~i=1,~ 2,~ 3, ~~~\alpha= 4, \cdots, 9 \nn\eea
where $A_i$, $A_\alpha$ and $\theta$ are N-dimensional vectors. These off-diagonal modes denote the interactions between the fuzzy sphere and D0-brane. To substitute the equation (\ref{fluctuation}) into the potential energy (\ref{pot}), we will get the terms of the fluctuational modes to second order.

First, we consider the bosonic parts of the potential (\ref{pot}). The first term is
\be -\frac{T_0}{4\lambda^2}\sum_{a\neq b} \Tr[X^a,
X^b]^2 = -\frac{T_0}{4\lambda^2}\Tr\left(\sum_{i\neq j}[X^i,
X^j]^2 + 2\sum_{i, \alpha} [X^i,
X^\alpha]^2+ \sum_{\alpha\neq \beta}[X^\alpha,
X^\beta]^2  \right). \ee With the equation (\ref{fluctuation}), all three terms in the above equation are
\be \Tr\left(\sum_{i\neq j}[X^i,
X^j]^2\right) = f^4\Tr([J_i, J_j]^2)+
4A_j^\dagger[K_i, K_j]A_i-4A_j^\dagger K_i^2A_j+4A_j^\dagger K_i K_jA_i, \ee
\be \Tr\left(\sum_{i, \alpha}[X^i,
X^\alpha]^2\right) = -2 A_\alpha^\dagger K_i^2A_\alpha,\ee
\be \Tr\left(\sum_{\alpha\neq \beta}[X^\alpha,
X^\beta]^2  \right)=\Tr\begin{pmatrix}
A_\alpha A_\beta^\dagger-A_\beta A_\alpha^\dagger & 0 \\
0 & A_\alpha^\dagger A_\beta-A_\beta^\dagger A_\alpha
\end{pmatrix}^2\ee where $K_i=fJ_i-x_i$.
To second order, the fluctuation of the first bosonic term is
\bea &&-\frac{T_0}{4\lambda^2}\sum_{a\neq b} \Tr[X^a,
X^b]^2= -\frac{T_0f^4}{4\lambda^2}\Tr([J_i, J_j]^2)\cr
&&~~~~~~~~~~~~-\frac{T_0}{\lambda^2}\left(A_j^\dagger[K_i, K_j]A_i-A_j^\dagger K_i^2A_j+A_j^\dagger K_i K_jA_i-A_\alpha^\dagger K_i^2A_\alpha\right).\label{bos1}\eea
Following the same way, the fluctuation of the second term in the potential energy is
\bea -\frac{i
  T_0}{3\lambda}\Tr(X^iX^jX^k)F^{(4)}_{0ijk}&=& \frac{i
  T_0f}{3\lambda^2}\Tr([X^i,X^j]X^k)\epsilon_{ijk}\cr
  &=&\frac{T_0f^4}{3\lambda^2}\Tr([J_i, J_j]^2)+\frac{2 i T_0f}{\lambda^2}\epsilon_{ijk}A_j^\dagger K_k A_i.\label{bos2}\eea
After combining the equation (\ref{bos1}) and (\ref{bos2}) together, we get the fluctuational modes of bosonic potential
\bea  \mathcal{V}_B
&=&-\frac{T_0 f^4}{24\lambda^2}N(N^2-1)\cr
&&+\frac{T_0}{\lambda^2}\left[A_j^\dagger(2if\epsilon_{ijk} K_k - [K_i, K_j])A_i+A_j^\dagger K_i^2A_j-A_j^\dagger K_i K_jA_i+A_\alpha^\dagger K_i^2A_\alpha\right]. \label{bpotential}\eea
Then we consider the fermionic part of the potential (\ref{pot}). The fluctuation at second order is
 \bea  \mathcal{V}_F&=&-\frac{T_0}{\lambda}\sum_a\Tr(\Theta^T\gamma_a[X^a,
\Theta])\cr
&=& -\frac{T_0}{\lambda}\theta^T\gamma_i K_i \theta. \label{fpotential}\eea

Now we get  the fluctuation of bosonic potential (\ref{bpotential}) and the fluctuation of the fermionic potential (\ref{fpotential}). Then we combine the two fluctuations together. Finally, the fluctuational potential to second order is  \be \mathcal{V}=\mathcal{V}_B+\mathcal{V}_F=\mathcal{V}^{(0)}+T_0 (A^\dagger\mathcal{M}_B^2 A-\theta^{T}\mathcal{M}_F\theta), \label{bfpotential}\ee with
\bea
\mathcal{V}^{(0)}&=&-\frac{T_0 f^4}{24\lambda^2}N(N^2-1),\cr
 A^\dagger\mathcal{M}_B^2 A&=&\frac{1}{\lambda^2}\left[A_j^\dagger(2if\epsilon_{ijk} K_k - [K_i, K_j])A_i+A_j^\dagger K_i^2A_j-A_j^\dagger K_i K_jA_i+A_\alpha^\dagger K_i^2A_\alpha\right],\cr
\theta^{T}\mathcal{M}_F\theta&=&\frac{1}{\lambda}\theta^T\gamma_i K_i \theta \eea where $\mathcal{M}_B^2$ and $\mathcal{M}_F$ are the mass matrices of fluctuational modes.

\subsection{Mass spectrum}
To consider the $N=2j+1$-dimensional irreducible representation of $SU(2)$, we choose the eigenvector $|j,m\rangle$ of operators $(J^2, J_3)$ to satisfy the equations
\be J^2|j,m\rangle=j(j+1)|j,m\rangle,~~~ J_3|j,m\rangle=m|j,m\rangle.\ee
We define $J_{\pm}=J_1\pm iJ_2$, $J_3^\dagger =J_3$ and $J_{\pm}^\dagger=J_{\mp}$.
The commutation relations about the operators $J_3$ and $J_{\pm}$ are
\be [J_3, J_{\pm}]=\pm J_{\pm},~~~[J_+, J_-]=2 J_3.\ee

The $J_{\pm}$ operators act on the eigenvector $|j,m\rangle$ as follows
\bea && J_+|j,m\rangle=\sqrt{(j-m)(j+m+1)}|j,m+1\rangle, \cr && J_-|j,m\rangle=\sqrt{(j+m)(j-m+1)}|j,m-1\rangle.\label{jpm}\eea
Then the off-diagonal fluctuations are expanded under this basis $|j, m\rangle$ as
\bea && A_a=\sum_{-j\le m\le j} \mathcal{A}_{a,m} |j,m\rangle,~~~A^\dagger_a=\sum_{-j\le m\le j} \langle j,m|\mathcal{A}^*_{a,m},\cr
 && \theta=\sum_{-j\le m\le j}\psi_m |j,m\rangle. \eea

Since the probe D0-brane is located at $x^1=x^2=0$ and $x^3=x=R+\Delta x$, we get $K_1=fJ_1$, $K_2=fJ_2$ and $K_3=fJ_3-x$. Then we have \be K_{\pm}=K_1\pm i K_2=f J_\pm,~~ K^2=\sum_{i=1}^3K_i^2=f^2J^2-2fxJ_3+x^2. \ee
By using the equation (\ref{jpm}), the operators $K^2$, $K_3$ and $K_{\pm}$ acting on the basis $|j,m\rangle$ will be
\bea  && K^2|j,m\rangle=\left[j(j+1)f^2-2mfx+x^2\right]|j,m\rangle,\cr
&& K_3|j,m\rangle=(fm-x)|j,m\rangle, \cr
&& K_{\pm}|j,m\rangle=f\sqrt{(j\mp m)(j\pm m+1)} |j,m\pm1\rangle. \label{kvalue}\eea
Obviously, the operators $K^2$ and $K_3$ are diagonal under the basis $|j,m\rangle$.

From the equation (\ref{bfpotential}), we get the mass matrix of the fluctuation $A_\alpha$, which is $K_i^2/\lambda^2$. Then, through using the equation (\ref{kvalue}), its eigenvalues are \be \omega^2_m(A_\alpha)=\frac{1}{\lambda^2}\left[j(j+1)f^2-2mfx+x^2\right], ~~-j\le m\le j. \label{massA}\ee

For the fluctuational fields $A_i$, their mass matrix can also be read off from the equation (\ref{bfpotential}). It is
\be M_B^2
= \frac{1}{\lambda^2}\begin{pmatrix} K^2-K_1^2 &-K_1K_2+2ifx &-K_1K_3 \\
-K_2K_1-2ifx &K^2-K_2^2 &-K_2K_3 \\
-K_3K_1 &-K_3K_2 & K^2-K_3^2  \end{pmatrix}\ee
In order to diagonalize this mass matrix, we choose a unitary matrix \be U=\begin{pmatrix} \frac{1}{\sqrt{2}} &-\frac{1}{\sqrt{2}} &0 \\ \frac{i}{\sqrt{2}} &\frac{i}{\sqrt{2}} &0\\
0 &0 &1\end{pmatrix},~~~~~~~ U^\dagger=\begin{pmatrix} \frac{1}{\sqrt{2}} &-\frac{i}{\sqrt{2}} &0 \\ -\frac{1}{\sqrt{2}} &-\frac{i}{\sqrt{2}} &0\\
0 &0 &1\end{pmatrix},~~~~U^\dagger U=1.\ee
Under the following unitary transformation, we get
\bea U^\dagger M_B^2 U&=& \frac{j(j+1)f^2+x^2}{\lambda^2}-\frac{1}{\lambda^2}N_B^2,\cr
N_B^2&=&\begin{pmatrix}\frac{1}{2}K_-K_++2fx(J_3+1) &-\frac{1}{2}K^2_- &\frac{1}{\sqrt{2}}K_-K_3 \\ -\frac{1}{2}K^2_+ &\frac{1}{2}K_+K_-+2fx(J_3-1) &-\frac{1}{\sqrt{2}}K_+K_3 \\
\frac{1}{\sqrt{2}}K_3K_+ &-\frac{1}{\sqrt{2}}K_3K_- &K_3^2+2fxJ_3\end{pmatrix}\label{nb}\eea
We assume the operator $N_B^2$ to satisfy the following eigenvalue equation
 \be N_B^2 \begin{pmatrix} \alpha_1 |j,m\rangle \\ \alpha_2 |j,m+2\rangle \\ \alpha_3 |j,m+1\rangle
\end{pmatrix}=v^2\begin{pmatrix} \alpha_1 |j,m\rangle \\ \alpha_2 |j,m+2\rangle \\ \alpha_3 |j,m+1\rangle
\end{pmatrix} \ee where $v^2$ is the eigenvalue, and $\alpha_i$ are some constants.
To insert the operator $N_B^2$ into the above equation, the eigenvalue equation is \be \begin{pmatrix} p^2+2(m+1)fx &-pq &pr \\
-pq &q^2+2(m+1)fx &-qr\\
pr &-qr &r^2+2(m+1)fx
\end{pmatrix}\begin{pmatrix}\alpha_1\\ \alpha_2\\ \alpha_3\end{pmatrix}=v^2\begin{pmatrix}\alpha_1\\ \alpha_2\\ \alpha_3\end{pmatrix}\ee where \bea && p=\frac{f}{\sqrt{2}}\sqrt{(j-m)(j+m+1)},\cr
&& q=\frac{f}{\sqrt{2}}\sqrt{(j-m-1)(j+m+2)},\cr
&& r=(m+1)f-x.\label{pqr}\eea
So three eigenvalues are  \be v^2= \left\{\begin{aligned}& 2(m+1)fx,\\ &2(m+1)fx,\\ &x^2+j(j+1)f^2.\end{aligned}\right.\label{nu2}\ee Obviously, the first two fluctuations are degenerate.
Substituting the above equation into Eq. (\ref{nb}), we obtain the mass spectra of $A_i$ as \be \omega_m^2(A_i)= \left\{\begin{aligned}& \frac{j(j+1)f^2+x^2-2(m+1)fx}{\lambda^2},\\ &\frac{j(j+1)f^2+x^2-2(m+1)fx}{\lambda^2},\\ &0.\end{aligned}\right.\label{massAA}\ee
The first $2N$ modes have same mass, and the last are $N$ zero modes. These zero modes mean some symmetries of this system maybe be broken.

Now we consider the fermionic fluctuations. Its mass matrix is \be \mathcal{M}_F=\frac{1}{\lambda}\sum_{i=1}^3\gamma_iK_i=\frac{1}{\lambda}
\left[f(\gamma_1J_1+\gamma_2J_2+\gamma_3J_3)-x\gamma_3\right].\ee Then the square of this mass matrix is \be \mathcal{M}_F^2=\frac{1}{\lambda^2}\left[(J_1^2+J_2^2+J_3^2)f^2+x^2-2fxJ_3+f^2N_F\right] \label{mf2}\ee where \be N_F=i\gamma_1\gamma_2 J_3+i\gamma_3\gamma_1J_2+i\gamma_2\gamma_3J_1.\ee With the relation $\{\gamma_i, \gamma_j\}=2\delta_{ij}$, we get
\be N_F^2=J^2+N_F.\ee If $u$ is the eigenvalue of the operator $N_F$, then it satisfies the equation \be u^2=j(j+1)+u.\ee  Thus, there are two eigenvalues $j+1$ and $-j$ for the operator $N_F$. We substitute these two eigenvalues into the equation (\ref{mf2}). Finally, the squares of mass of $8N$ fermionic modes are \bea \omega_m^2(\theta_+)&=&\frac{1}{\lambda^2}\left[(j+1)^2f^2+x^2-2mfx\right],\cr
\omega_m^2(\theta_-)&=&\frac{1}{\lambda^2}\left[j^2f^2+x^2-2mfx\right].\label{massf}\eea
Each one is a $4N$-fold degeneracy.

\section{Entropic force}
Now we already found there are $6N$ $\omega_m(A_\alpha)$ and $2N$  $\omega_m(A_i)$ bosonic modes. Also there are  $4N$  $\omega_m(\theta_+)$ and $4N$  $\omega_m(\theta_-)$ fermionic modes. If we take the time coordinate to be periodic, then the fluctuations on the fuzzy sphere will be a thermodynamic system with a finite temperature $T$. After including all the contributions of the bosonic and fermionic modes, we obtain the partition function of this system, which is \bea \mathcal{Z}&=&\prod_{\rm B}\frac{1}{1-e^{-\beta \omega_B}}\prod_{\rm F}(1+e^{-\beta\omega_F})\cr
&=& \prod_{m=-j}^j\left[\frac{1}{1-e^{-\beta\omega_m(A_\alpha)}}\right]^{6(2j+1)}\left[\frac{1}{1-e^{-\beta\omega_m(A_i)}}\right]^{2(2j+1)}\cr
&&\times\prod_{m=-j}^j\left[1+e^{-\beta\omega_m(\theta_+)}\right]^{4(2j+1)}\left[1+e^{-\beta\omega_m(\theta_-)}\right]^{4(2j+1)}
\label{partition}\eea where the indices $\rm B$ and $\rm F$ denote all bosonic and fermionic modes respectively. Then, the free energy of this system is \be {\mathcal F}=-\beta^{-1}\ln \mathcal{Z}. \label{free1}\ee
We insert the equation (\ref{partition}) into (\ref{free1}), the free energy is \bea  {\mathcal F}&=&\beta^{-1}(2j+1)\sum_{m=-j}^j\left[6\ln\left(1-e^{-\beta\omega_m(A_\alpha)}\right)+
2\ln\left(1-e^{-\beta\omega_m(A_i)}\right)\right. \cr&& \left. -4\ln\left(1+e^{-\beta\omega_m(\theta_+)}\right)
-4\ln\left(1+e^{-\beta\omega_m(\theta_-)}\right)\right]\cr
&=&\beta^{-1}(2j+1)\left[6\mathcal{J}_+(A_\alpha)+2\mathcal{J}_+(A_i)
-4\mathcal{J}_-(\theta_+)-4\mathcal{J}_-(\theta_-)\right] \label{calI}
\eea
where \be \mathcal{J}_{\pm}(\phi)=\sum_{m=-j}^j\ln(1\mp e^{-\beta \omega_m(\phi)})\ee and the field $\phi$ denotes the modes $A_\alpha$, $A_i$, $\theta_+$ and $\theta_-$.
Approximate to the first order of $dx$, the mass of modes is  \bea &&\omega_m(\phi)=\omega_m^{(0)}(\phi)+\omega_m^{(1)}(\phi)\cdot dx,\cr
&&\omega_m^{(0)}(\phi)=\omega_m(\phi)\big|_{x=R},~~~~\omega_m^{(1)}(\phi)=\frac{\pt\omega_m(\phi)}{\pt x}\bigg|_{x=R}.\eea Then, we get
\bea \mathcal{J}_{\pm}(\phi)&\approx &\sum_{m=-j}^j\ln\left[1\mp e^{-\beta\left(\omega_m^{(0)}(\phi)+\omega_m^{(1)}(\phi)\cdot dx\right) }\right]\cr
&\approx& \sum_{m=-j}^j\ln\left[1\mp e^{-\beta\omega_m^{(0)}(\phi)}\right]+
\sum_{m=-j}^j\ln\left[1\pm\frac{\beta\omega_m^{(1)}(\phi)}{e^{\beta\omega_m^{(0)}(\phi)}\mp 1}dx\right] \cr
&=& \mathcal{J}_{\pm}^{(0)}(\phi)\pm \mathcal{J}_{\pm}^{(1)}(\phi)\cdot dx\label{jpn}\eea with the definitions \be \mathcal{J}_{\pm}^{(0)}(\phi)= \sum_{m=-j}^j\ln\left[1\mp e^{-\beta\omega_m^{(0)}(\phi)}\right],~~~\mathcal{J}_{\pm}^{(1)}(\phi)=\sum_{m=-j}^j\frac{\beta\omega_m^{(1)}(\phi)}{e^{\beta\omega_m^{(0)}(\phi)}\mp 1}. \label{j0j1}\ee

The free energy can be expanded as \be {\mathcal F}={\mathcal F}^{(0)}+{\mathcal F}^{(1)}\cdot dx +\cdots.\label{free2}\ee By substituting the equation (\ref{jpn}) into the equation (\ref{calI}), we obtain \bea {\mathcal F}^{(0)}&=&\beta^{-1}(2j+1)\left[6\mathcal{J}^{(0)}_+(A_\alpha)+2\mathcal{J}^{(0)}_+(A_i)
-4\mathcal{J}^{(0)}_-(\theta_+)-4\mathcal{J}^{(0)}_-(\theta_-)\right],\cr
{\mathcal F}^{(1)}&=&\beta^{-1}(2j+1)\left[6\mathcal{J}^{(1)}_+(A_\alpha)+2\mathcal{J}^{(1)}_+(A_i)
+4\mathcal{J}^{(1)}_-(\theta_+)+4\mathcal{J}^{(1)}_-(\theta_-)\right]. \label{free}\eea
Following the equations (\ref{massA}), (\ref{massAA}) and (\ref{massf}), we get the mass of modes at zero order of $dx$
\bea \omega^{(0)}_m(A_\alpha)&=&
\frac{\sqrt{2}R}{\lambda}\sqrt{1-\frac{m}{\sqrt{j(j+1)}}},\cr
\omega^{(0)}_m(A_i)&=&
\frac{\sqrt{2}R}{\lambda}\sqrt{1-\frac{m+1}{\sqrt{j(j+1)}}},\cr
\omega^{(0)}_m(\theta_+)&=&
\frac{\sqrt{2}R}{\lambda}\sqrt{1+\frac{1}{2j}-\frac{m}{\sqrt{j(j+1)}}},\cr
\omega^{(0)}_m(\theta_-)&=&
\frac{\sqrt{2}R}{\lambda}\sqrt{1-\frac{1}{2(j+1)}-\frac{m}{\sqrt{j(j+1)}}},
\eea where the radius of the fuzzy sphere is $R=\frac{f\sqrt{N^2-1}}{2}=\sqrt{j(j+1)}f$.
And the mass of modes at first order of $dx$ reads
\bea
\omega^{(1)}_m(A_\alpha)&=&\frac{1}{\sqrt{2}\lambda}\sqrt{1-\frac{m}{\sqrt{j(j+1)}}},\cr
\omega^{(1)}_m(A_i)&=&\frac{1}{\sqrt{2}\lambda}\sqrt{1-\frac{m+1}{\sqrt{j(j+1)}}},\cr
\omega^{(1)}_m(\theta_+)&=&\frac{1}{\sqrt{2}\lambda}\frac{1-\frac{m}{\sqrt{j(j+1)}}}{\sqrt{1+\frac{1}{2j}-\frac{m}{\sqrt{j(j+1)}}}},\cr
\omega^{(1)}_m(\theta_-)&=&\frac{1}{\sqrt{2}\lambda}\frac{1-\frac{m}{\sqrt{j(j+1)}}}{\sqrt{1-\frac{1}{2(j+1)}-\frac{m}{\sqrt{j(j+1)}}}}.
\eea
In the large $N$ limit, the mass of modes at zero and first order of $dx$ are
\be \omega^{(0)}_m(\phi)\approx\frac{\sqrt{2}R}{\lambda}\sqrt{1-\frac{m}{j}},
~~~~~\omega^{(1)}_m(\phi)\approx \frac{1}{\sqrt{2}\lambda}\sqrt{1-\frac{m}{j}}.\ee
Clearly, the mass of all modes at the zero and first order of $dx$ are the same. Then the free energy in the equation (\ref{free}) in the large $N$ limit reduces to
\bea {\mathcal F}^{(0)}&=&8\beta^{-1}(2j+1)\left[\mathcal{J}^{(0)}_+(\phi)
-\mathcal{J}^{(0)}_-(\phi)\right], \cr
{\mathcal F}^{(1)}&=&8\beta^{-1}(2j+1)\left[\mathcal{J}^{(1)}_+(\phi)+\mathcal{J}^{(1)}_-(\phi)\right] \label{free11}\eea where $\mathcal{J}_{\pm}^{(0)}(\phi)$ and $\mathcal{J}_{\pm}^{(1)}(\phi)$ in the equation (\ref{j0j1}) become
\bea \mathcal{J}^{(0)}_{\pm}(\phi)&=& \sum_{m=-j}^j\ln\left[1\mp e^{-\frac{\sqrt{2}\beta R}{\lambda}\sqrt{1-\frac{m}{j}}}\right],\cr
\mathcal{J}^{(1)}_{\pm}(\phi)&=&\sum_{m=-j}^j\frac{\frac{\beta}{\sqrt{2}\lambda}\sqrt{1-\frac{m}{j}}}{e^{\frac{\sqrt{2}\beta R}{\lambda}\sqrt{1-\frac{m}{j}}}\mp 1}.\label{j1pn}\eea

From the equation (\ref{free2}), the entropy is \be S=-\frac{\pt {\mathcal F}}{\pt T}\approx k_B\beta^2\left(\frac{\pt {\mathcal F}^{(0)}}{\pt\beta}+\frac{\pt {\mathcal F}^{(1)}}{\pt\beta}\cdot dx \right)=S^{(0)}+\frac{\pt S}{\pt x}\cdot dx\ee where \be S^{(0)}=k_B\beta^2\frac{\pt {\mathcal F}^{(0)}}{\pt\beta},~~~\frac{\pt S}{\pt x}=k_B\beta^2\frac{\pt {\mathcal F}^{(1)}}{\pt\beta}\ee  In the above equation, $S^{(0)}$ is the entropy at zero order of $d x$. Following the definition of entropic force in \cite{Verlinde:2010hp}, we get \be F=T\frac{\pt S}{\pt x}=\beta\frac{\pt {\mathcal F}^{(1)}}{\pt \beta}. \ee By using the equations (\ref{free11}) and (\ref{j1pn}), the entropy at zero order and entropic force are  \bea
S^{(0)}&=&-8k_BN\sum_{m=-j}^{m=j}\left[\ln\frac{1-e^{-\frac{\sqrt{2}\beta R}{\lambda}\sqrt{1-\frac{m}{j}}}}{1+e^{-\frac{\sqrt{2}\beta R}{\lambda}\sqrt{1-\frac{m}{j}}}}\right]+\cr &&8k_BN\beta\sum_{m=-j}^{m=j}
\left[\frac{\frac{\sqrt{2}R}{\lambda}\sqrt{1-\frac{m}{j}}e^{-\frac{\sqrt{2}\beta R}{\lambda}\sqrt{1-\frac{m}{j}}}}{1-e^{-\frac{\sqrt{2}\beta R}{\lambda}\sqrt{1-\frac{m}{j}}}}+\frac{\frac{\sqrt{2}R}{\lambda}\sqrt{1-\frac{m}{j}}e^{-\frac{\sqrt{2}\beta R}{\lambda}\sqrt{1-\frac{m}{j}}}}{1+e^{-\frac{\sqrt{2}\beta R}{\lambda}\sqrt{1-\frac{m}{j}}}}\right]\\
F&=&-\frac{8N\beta R}{\lambda^2}\sum_{m=-j}^{m=j}\left[\frac{(1-\frac{m}{j})e^{\frac{\sqrt{2}\beta R}{\lambda}\sqrt{1-\frac{m}{j}}}}{(e^{\frac{\sqrt{2}\beta R}{\lambda}\sqrt{1-\frac{m}{j}}}+1)^2}+\frac{(1-\frac{m}{j})e^{\frac{\sqrt{2}\beta R}{\lambda}\sqrt{1-\frac{m}{j}}}}{(e^{\frac{\sqrt{2}\beta R}{\lambda}\sqrt{1-\frac{m}{j}}}-1)^2}\right].\eea
At large $N$, $m/j$ can be taken as a continuous variable $y$. Defining a dimensionless variable $\tau=\frac{\sqrt{2}\beta R}{\lambda}\sqrt{1-y}$, we get \bea
S^{(0)}&=&\frac{4k_BN(N-1)\lambda^2}{\beta^2R^2}\int_0^{\frac{2\beta R}{\lambda}}d\tau\left[\frac{2\tau^2e^{-\tau}}{1-e^{-2\tau}}+\tau\ln\frac{1+e^{-\tau}}{1-e^{-\tau}}\right]
, \cr
F&=&-\frac{4N(N-1)\lambda^2}{\beta^3R^3}\int_0^{\frac{2\beta R}{\lambda}}d\tau \frac{\tau^3 e^\tau(e^{2\tau}+1)}{(e^{2\tau}-1)^2}. \label{force}\eea

Now we will analyze the integrations (\ref{force}). First, we consider the large limit of the radius $R\propto N f$ with the large $N$ and finite four-form flux $f$. Then the entropy at zero order of $dx$ and entropic force are \be S^{(0)}=\frac{21\zeta(3)k_BN^2\lambda^2}{\beta^2R^2},~~~~F=-\frac{21\zeta(3)N^2\lambda^2}{\beta^3R^3}.\ee
We let the time direction to be periodic with a radius $R$. Then the temperature of this thermal dynamic system is taken to be \be T\propto \frac{1}{R}.\ee  Thus, we get \be S^{(0)}\propto\frac{1}{N^2}, ~~~F\propto -\frac{1}{R^6}.\ee  Obviously, this result does not go like the case in \cite{Verlinde:2010hp}. The fuzzy sphere can not be taken as a holographic screen. The reason is that the commutation $[X^i, X^j]$ is not commutative due to the nonvanishing four-form flux $f$.

Finally, we consider the second limit. We choose the four-form flux $f$ as infinitely small and $N$ as still large, then the radius $R$ remains infinitely small. In this case, the RR four-form is almost vanished. Then the commutation relation about the coordinates $X^i$ satisfies $[X^i, X^j]\approx 0$. It means the fuzzy sphere almost becomes a usual sphere. From the equations (\ref{force}), we get \bea S^{(0)}&\approx& \frac{6k_BN^2\lambda^2}{\beta^2 R^2}\int_0^{\frac{2\beta R}{\lambda}}d\tau\propto N^2,\\
F&\approx& -\frac{6N^2\lambda^2}{\beta^3 R^3}\int_0^{\frac{2\beta R}{\lambda}}\tau d\tau\propto-\frac{1}{\beta R}\propto -\frac{1}{R^2}.\eea The entropy $S^{(0)}$ is  proportional to the area $4\pi R^2$ of the fuzzy sphere. So it is really a holographic screen, which is composed of $N$ D0-branes with the four-form flux $f\approx 0$. And the entropic force is proportional to $1/R^2$. The minus means the force is along the negative direction of the $x^3$ axis. At small radius, the Yukawa-type high dimensional correction to the gravitational inverse square law is order one in \cite{Kehagias:1999my}. Thus, the force of gravitation in bulk is entropic in the very small radius $R$ and RR four-form regime.

Now we give some discussion about our result. The supersymmetry is broken at finite temperature. Then the high order corrections of bosonic and fermionic modes are not under control. But these high order perturbative corrections will not change our result in the small $R$ and zero four-form RR flux $f$ limit. When a probe D0-brane approaches this fuzzy sphere and is finally dissolved, the distance between D0-branes on the fuzzy sphere is very small in this limit. Thus, the off-diagonal fluctuations become very small. Thus, the higher order corrections still can be ignored, and our perturbative calculation is reliable.

\subsection*{Acknowledgments}

We are very glad to thank Professor Yi-hong Gao for the
collaboration in this project. We also thank Professor Zhu-feng Zhang for very helpful discussions. This work is partly supported by the K. C. Wong Magna Fund in Ningbo University.

\appendix

\end{document}